\documentclass[showpacs,amsmath,amsart,twocolumn,showkeys,pra,superscriptaddress]{revtex4-1}
\usepackage{dcolumn}    
\usepackage{ifpdf}
\usepackage{amssymb,lineno,amsfonts}
\usepackage{graphicx}   
\usepackage{bm}         
\usepackage{bbm}
\usepackage{mathrsfs}
\usepackage{upgreek}
\usepackage{mathtools}
\usepackage{epstopdf}
\usepackage{setspace}
\usepackage{hyperref}
\usepackage{float}
\usepackage{natbib}
\usepackage[usenames,dvipsnames]{xcolor}
\usepackage[matrix,frame,arrow]{xypic}

\newcommand{\ket}[1]{\vert{#1}\rangle}
\newcommand{\bra}[1]{\langle{#1}\vert}
\newcommand{\outpr}[2]{\vert{#1}\rangle\langle{#2}\vert}

\newcommand{\expec}[1]{\langle{#1}\rangle}
\definecolor{med-blue}{RGB}{25,25,112}
\hypersetup{colorlinks, linkcolor={red},citecolor={blue}, urlcolor={MidnightBlue}}

\begin{document}
\title{Violation of space-time Bell-CHSH inequality beyond Tsirelson bound via post selection and quantum cryptography}
\author{C S Sudheer Kumar} 
\email{sudheer.kumar@students.iiserpune.ac.in}
\affiliation{NMR Research Center and Department of Physics, \\ Indian Institute of Science Education and Research, Pune 411008, India} 
\begin{abstract}
Here we show that, if we insert context dependent unitary evolutions (which can be achieved via post selection) into spatial (i.e., normal) Bell-CHSH test, then it is possible to violate space-time Bell-CHSH inequality maximally (i.e., up to $4$). 
However this does not contradict Tsirelson quantum bound ($2\sqrt{2}$), as the latter has been derived without taking into consideration context dependent unitary evolutions and/or post selection. As an important application, this leads to a more efficient quantum key distribution protocol.
\end{abstract}
\keywords{Space-time Bell-CHSH test, quantum key distribution}
\pacs{03.65.Ta, 03.67.Dd, 03.65.Ud, 03.67.-a}
\maketitle

``Correlations cry out for explanation" -J S Bell \cite{Bell_spek_unspek_book}. 
It has been already shown that, it is possible to violate Bell-CHSH inequality maximally (i.e., up to $4$) without violating special relativity \cite{Bellvio4Popescu1994,Bellvio4postselectcabPRL}. Here we present yet another method of achieving the same in a nontrivial fashion i.e., via context dependent unitary evolutions (which can be achieved through post selection), without the need for any type of signalling. In the spatial (i.e., normal) Bell-CHSH test, there is no unitary evolution. Alice and Bob randomly choose their observables and directly measure them locally on their respective entangled qubit state. As entangled particles are correlated over space in spite of measurement events being space-like separated (nonlocal correlation), correlation between Alice's and Bob's observables can go up to $2\sqrt{2}$, there by violating local realistic bound, $2$. It is possible to boost the correlation over space (which resulted in $2\sqrt{2}$) further via context dependent unitary evolutions. But this requires Bob to know what Alice has measured, which is not possible unless Alice can signal Bob. Hence Bob applies unitary operations randomly and then measures. Later they post select correct context dependent unitarily evolved states. By this Alice and Bob can achieve maximum possible correlation (i.e., $4$) between their observables. However, this does not contradict Tsirelson quantum bound ($2\sqrt{2}$), as the latter has been derived without taking into consideration context dependent unitary evolutions and/or post selection \cite{Tsirelsonboundoriginalpaper,quant_info_neilson_chuang}.

As an important application, this leads to a more efficient quantum key distribution (QKD) protocol. QKD or quantum cryptography is a provably secure protocol using which private key bits can be generated between two parties over a public channel \cite{quant_info_neilson_chuang}. Security of QKD protocols is based on the fact that, eavesdropper cannot steal the information without disturbing the quantum state. Suppose Alice want to send Bob a secret message `Hi'. They some how have shared a secret key `qw' (e.g., they met personally in the past and shared (but this is not always feasible) or via QKD). Alice mixes her secret message with the secret key (encryption) and obtains `Hi+qw=rd'. Alice sends `rd' to Bob over public channel. Then Bob decrypts the message to retrieve original secret message: `rd-qw=Hi'. Mainly there are two types of QKD protocols: (1) BB84 (not based on Bell's theorem) \cite{quant_info_neilson_chuang,QKD_expt_entphtn_Zeilinger}. (2) Ekert's QKD protocol which uses Bell's theorem for its security \cite{QKD_Ekert_Bell, Quant_crypt_review_Gisin}. In our QKD protocol approximately half of the resource (which correspond to correct context dependent unitarily evolved states) is utilised for secret key bits generation, and approximately half of the resource (which correspond to wrong context dependent unitarily evolved states) along with a small amount of correct context dependent unitarily evolved states, is utilised to test for eavesdropping. Hence our QKD protocol utilises the resource more efficiently than other QKD protocols, there by enhancing the total amount of key bits generated and also the security. QKD has become important, because the security of public key distribution protocols is under threat with the advent of quantum computers, which can find the prime factors of large numbers in polynomial time (Shor's algorithm) \cite{quant_info_neilson_chuang}.

\textit{Space-time Bell-CHSH test} : Alice and Bob share $N$ number of singlets: $\ket{S_0}=(\ket{01}-\ket{10})/\sqrt{2}=-(\ket{+-}-\ket{-+})/\sqrt{2}$ where $\ket{0},\ket{1}$ are eigenkets of $\sigma_z$ (Pauli-z matrix) with eigenvalues $+1,-1$ respectively, and $\ket{\pm}=(\ket{0}\pm\ket{1})/\sqrt{2}$. Alice measures locally the observables $A=\sigma_z\otimes\mathbbm{1},C=\sigma_x\otimes\mathbbm{1}$ on her qubits, with probability $1/2,1/2$ respectively. Bob measures locally the observables $B=-\mathbbm{1}\otimes(\sigma_z+\sigma_x)/\sqrt{2},D=\mathbbm{1}\otimes(\sigma_z-\sigma_x)/\sqrt{2}$ on his qubits, with probability $1/2,1/2$ respectively where $\sigma_{i}$ is Pauli-$i$ matrix, $i=x,y,z$, and $\mathbbm{1}$ is $2\times 2$ identity matrix \cite{quant_info_neilson_chuang}. Alice's and Bob's clocks are synchronised and their measurement events are space-like separated. Alice measures her $j^{\mbox{th}}$ qubit state at time $t=t^A_j$. Immediately after Alice's measurement, Bob applies unitary operator $U_k$ to his $j^{\mbox{th}}$ qubit state and then measures at time $t=t^B_j(>t^A_j)$, where $U_k$ is chosen randomly from the set $\{U_{+y},U_{-y}\}$ with probability $\{1/2,1/2\}$ respectively, $j=1,2,...,N$. Bob knows each of $t^A_j$s. As collapse is instantaneous (which is evident from violation of spatial Bell-CHSH inequality \cite{Bell_vio_lopholefree_1pt3km}), Bob can carry out his operations immediately after Alice measures (hence Bob need not have to store his qubit states in quantum memory). We have the following eigenvalue equations:
\begin{eqnarray}
\sigma_z\ket{0}&=&\ket{0},\sigma_z\ket{1}=-\ket{1},\sigma_x\ket{\pm}=\pm\ket{\pm},\nonumber\\
\frac{-(\sigma_z+\sigma_x)}{\sqrt{2}}\ket{\pm}_B&=&\pm\ket{\pm}_B,\frac{\sigma_z-\sigma_x}{\sqrt{2}}\ket{\pm}_D=\pm\ket{\pm}_D~~~~~
\end{eqnarray}
where 
\begin{eqnarray}
\ket{+}_B&=&\cos(\theta_1/2)\ket{0}+e^{i\pi}\sin(\theta_1/2)\ket{1},\nonumber\\
\ket{-}_B&=&\cos(\theta_2/2)\ket{0}+\sin(\theta_2/2)\ket{1},\nonumber\\
\ket{+}_D&=&\cos(\theta_2/2)\ket{0}+e^{i\pi}\sin(\theta_2/2)\ket{1},\nonumber\\
\ket{-}_D&=&\cos(\theta_1/2)\ket{0}+\sin(\theta_1/2)\ket{1},
\end{eqnarray}
$\theta_1=\pi-\pi/4,\theta_2=\pi/4$. Quantum mechanically the values of measurement outcomes $a,c,b,d$ ($=\pm 1$, the eigenvalues) of observables $A,C,B,D$ respectively, are not preassigned before the measurement process. $b,d$ depends on Alice's choice of observable, even though their measurement events are space-like separated. Measurement creates reality. \\When Alice measures $A$ locally, if her qubit collapses to $\ket{0}$ or $\ket{1}$, then Bob's qubit always collapses instantaneously to $\ket{1}$ or $\ket{0}$ respectively (spatial correlation due to entanglement). Similarly when Alice measures $C$ locally, if her qubit collapses to $\ket{\pm}$, then Bob's qubit collapses to $\ket{\mp}$. 

After $N$ measurements, they select out (via classical communication) the following four (out of eight) subensembles which correspond to correct context dependent unitary evolutions:

($\mathcal{E}_1$) Alice had measured $A$, then Bob had evolved his qubit state under the unitary $U_{+y}=\exp(-i(\pi/4)(\sigma_y/2))$ (i.e., counter clock wise rotation about y-axis by $45^o$ on the Bloch sphere) to get $U_{+y}\ket{1}=\ket{+}_B$ or $U_{+y}\ket{0}=\ket{-}_B$, and then he had measured $B$. As $\ket{+}_B,\ket{-}_B$ are eigenkets of $B$, variance in Bob's measurement outcome is exactly zero. Then the product of measurement outcomes becomes $ab=+1\times +1=1$ or $ab=-1\times -1=1$. Hence knowing $b$, Bob can know $a$ i.e., $a=b$ (perfectly correlated). Hence $\expec{A(t^A)B_1(t^B)}=1=ab$ where $B_i,D_i$ represents association of $B,D$ respectively with unitary evolution $U_{a_i},i=1,2,a_1=+y,a_2=-y$. More rigorously, joint probability of Alice getting outcome $a$ in a measurement of $A$ and Bob, after applying $U_{+y}$, getting outcome $b$ in a measurement of $B$ is given by
\begin{eqnarray}
p(a,b)=\mbox{Tr}(\mathcal{B}^{(2)}_b U^{(2)}_{+y}\mathcal{A}^{(1)}_a\rho_0\mathcal{A}^{(1)}_a U^{(2)\dagger}_{+y})
\label{pab}
\end{eqnarray}   
where, $a,b=+1,-1$, $\rho_0=\outpr{S_0}{S_0}$, $\mathcal{A}^{(1)}_{+1}=\outpr{0}{0}\otimes\mathbbm{1},\mathcal{A}^{(1)}_{-1}=\outpr{1}{1}\otimes\mathbbm{1}$, $\mathcal{B}^{(2)}_{\pm 1}=\mathbbm{1}\otimes\ket{\pm}_B\bra{\pm}_B$, $U^{(2)}_{+y}=\mathbbm{1}\otimes U_{+y}$.  $\Rightarrow\expec{A(t^A)B_1(t^B)}=\sum_{a,b}p(a,b)~ab=1$, where $p(+1,+1)=p(-1,-1)=1/2$ and $p(+1,-1)=p(-1,+1)=0$. Or $\expec{A(t^A)B_1(t^B)}=\sum_{a,b}p(a,b)~ab=ab\sum_{a,b}p(a,b)=ab=1$ ($\because~ab=1$).

($\mathcal{E}_2$) Alice had measured $A$, then Bob had evolved his qubit state under the unitary $U_{-y}=\exp(i(\pi/4)(\sigma_y/2))$ (i.e., clock wise rotation about y-axis by $45^o$ on the Bloch sphere) to get $U_{-y}\ket{1}=\ket{-}_D$ or $U_{-y}\ket{0}=\ket{+}_D$, and then he had measured $D$. Again no variance in Bob's measurement outcome. Then the product of measurement outcomes becomes $ad=+1\times -1=-1$ or $ad=-1\times +1=-1$. $\Rightarrow a=-d$ (perfectly anticorrelated). Hence $\expec{A(t^A)D_2(t^B)}=-1=ad$. More rigorously, joint probability of Alice getting outcome $a$ in a measurement of $A$ and Bob, after applying $U_{-y}$, getting outcome $d$ in a measurement of $D$ is given by
\begin{eqnarray}
p(a,d)=\mbox{Tr}(\mathcal{D}^{(2)}_d U^{(2)}_{-y}\mathcal{A}^{(1)}_a\rho_0\mathcal{A}^{(1)}_a U^{(2)\dagger}_{-y})
\label{pad}
\end{eqnarray}   
where, $a,d=+1,-1$, $\mathcal{D}^{(2)}_{\pm 1}=\mathbbm{1}\otimes\ket{\pm}_D\bra{\pm}_D$, $U^{(2)}_{-y}=\mathbbm{1}\otimes U_{-y}$. $\Rightarrow\expec{A(t^A)D_2(t^B)}=\sum_{a,d}p(a,d)~ad=-1=ad$ where, $p(+1,+1)=p(-1,-1)=0$ and $p(+1,-1)=p(-1,+1)=1/2$. 

($\mathcal{E}_3$) Alice had measured $C$, then Bob had evolved his qubit state under the unitary $U_{-y}$ to get $U_{-y}\ket{-}=\ket{+}_B$ or $U_{-y}\ket{+}=\ket{-}_B$, and then he had measured $B$. Again no variance in Bob's measurement outcome. Then the product of measurement outcomes become $cb=+1\times +1=1$ or $cb=-1\times -1=1$. $\Rightarrow c=b$ (perfectly correlated). Hence $\expec{C(t^A)B_2(t^B)}=1=cb$. More rigorously, joint probability of Alice getting outcome $c$ in a measurement of $C$ and Bob, after applying $U_{-y}$, getting outcome $b$ in a measurement of $B$ is given by
\begin{eqnarray}
p(c,b)=\mbox{Tr}(\mathcal{B}^{(2)}_b U^{(2)}_{-y}\mathcal{C}^{(1)}_c\rho_0\mathcal{C}^{(1)}_c U^{(2)\dagger}_{-y})
\label{pcb}
\end{eqnarray}   
where, $c,b=+1,-1$, $\mathcal{C}^{(1)}_{\pm 1}=\outpr{\pm}{\pm}\otimes\mathbbm{1}$. $\Rightarrow \expec{C(t^A)B_2(t^B)}=\sum_{c,b}p(c,b)cb=1=cb$, where $p(+1,+1)=p(-1,-1)=1/2$ and $p(+1,-1)=p(-1,+1)=0$. 

($\mathcal{E}_4$) Alice had measured $C$, then Bob had evolved his qubit state under the unitary $U_{+y}$ to get $U_{+y}\ket{-}=\ket{+}_D$ or $U_{+y}\ket{+}=\ket{-}_D$, and then he had measured $D$. Again no variance in Bob's measurement outcome. Then the product of measurement outcomes becomes $cd=+1\times +1=1$ or $cd=-1\times -1=1$. $\Rightarrow c=d$ (perfectly correlated). Hence $\expec{C(t^A)D_1(t^B)}=1=cd$. More rigorously, joint probability of Alice getting outcome $c$ in a measurement of $C$ and Bob, after applying $U_{+y}$, getting outcome $d$ in a measurement of $D$ is given by
\begin{eqnarray}
p(c,d)=\mbox{Tr}(\mathcal{D}^{(2)}_d U^{(2)}_{+y}\mathcal{C}^{(1)}_c\rho_0\mathcal{C}^{(1)}_c U^{(2)\dagger}_{+y}),
\label{pcd}
\end{eqnarray}   
where $c,d=+1,-1$. $\Rightarrow \expec{C(t^A)D_1(t^B)}=\sum_{c,d}p(c,d)cd=1=cd$ where $p(+1,+1)=p(-1,-1)=1/2$ and $p(+1,-1)=p(-1,+1)=0$.  

Now substituting the expectation values into the space-time Bell-CHSH term we obtain
\begin{eqnarray}
\expec{I_{Q}}=\expec{A(t^A)B_1(t^B)}+\expec{C(t^A)B_2(t^B)}\nonumber\\
+\expec{C(t^A)D_1(t^B)}-\expec{A(t^A)D_2(t^B)}=4\nonumber\\
=ab+cb+cd-ad=I_Q,
\label{I_Q}
\end{eqnarray}
which is the maximum possible violation of classical (local) upper bound, $2$ (see Appendix (\ref{appdx_no_postsel}) for derivation of classical upper bound). $I_Q$ takes only one value i.e., $4$. Hence $\expec{I_Q}=I_Q=4$. Note that $ab+cb+cd-ad\ne (a+c)b-(a-c)d\le 2$ because $d$ in the context of $a$ is different from $d$ in the context of $c$. Similar thing with $b$. The fact that $\expec{I_Q}=4$ does not contradict Tsirelson bound ($2\sqrt{2}$) \cite{Tsirelsonboundoriginalpaper,quant_info_neilson_chuang}, because there is unitary evolution involved, and Alice and Bob are post selecting correct context dependent unitarily evolved subensembles. Both of these are not considered in deriving Tsirelson bound. 


There are two context dependencies here: (1) Whether Bob measures $B$ in the context of $A$ or in the context of $C$ ($A$ and $C$ do not commute). This context dependency manifests as nonlocal correlation over space as measurement events are space-like separated. Similar context dependency for $D$. This results in $2<\expec{I_{Q}}\le 2\sqrt{2}$. (2) The context dependent unitary operations that Bob applies to his qubit states, as described above. This boosts the nonlocal correlation over space that is already present, to the maximum extent possible. If there was no nonlocal correlation over space (like in classical scenario), then time evolution cannot boost the correlation any further. State of Bob's qubit gets maximally (anti)correlated (with respect to measurement outcomes) with that of Alice's, as Bob applies $U_{\pm y}$. This results in $2\sqrt{2}<\expec{I_{Q}}\le 4$. \\Time evolution induces/causes perfect synchronization of spin polarization directions of Alice's and Bob's qubit states. During time evolution, Bob's qubit evolves into such a state that measurement outcomes of Alice and Bob gets resonated.

Remaining four subensembles correspond to completely wrong context dependent unitarily evolved states (i.e., not even a single correct context dependent unitary evolution). They give following joint probabilities and expectation values:

($\mathcal{E}_5$)
\begin{eqnarray}
p(a,b)=\mbox{Tr}(\mathcal{B}^{(2)}_b U^{(2)}_{-y}\mathcal{A}^{(1)}_a\rho_0\mathcal{A}^{(1)}_a U^{(2)\dagger}_{-y}),
\label{pab1}
\end{eqnarray}   
where $a,b=+1,-1$. $\Rightarrow\expec{A(t^A)B_1(t^B)}=\sum_{a,b}p(a,b)~ab=0$, where $p(+1,+1)=p(-1,-1)=p(+1,-1)=p(-1,+1)=1/4$, and $B_i,D_i$ represents association of $B,D$ respectively with unitary evolution $U_{a_i},i=1,2,a_1=-y,a_2=+y$.

($\mathcal{E}_6$)
\begin{eqnarray}
p(a,d)=\mbox{Tr}(\mathcal{D}^{(2)}_d U^{(2)}_{+y}\mathcal{A}^{(1)}_a\rho_0\mathcal{A}^{(1)}_a U^{(2)\dagger}_{+y})
\label{pad1}
\end{eqnarray}   
where, $a,d=+1,-1$. $\Rightarrow\expec{A(t^A)D_2(t^B)}=\sum_{a,d}p(a,d)~ad=0$ where $p(+1,+1)=p(-1,-1)=p(+1,-1)=p(-1,+1)=1/4$. 

($\mathcal{E}_7$)
\begin{eqnarray}
p(c,b)=\mbox{Tr}(\mathcal{B}^{(2)}_b U^{(2)}_{+y}\mathcal{C}^{(1)}_c\rho_0\mathcal{C}^{(1)}_c U^{(2)\dagger}_{+y})
\label{pcb1}
\end{eqnarray}   
where, $c,b=+1,-1$. $\Rightarrow \expec{C(t^A)B_2(t^B)}=\sum_{c,b}p(c,b)cb=0$, where $p(+1,+1)=p(-1,-1)=p(+1,-1)=p(-1,+1)=1/4$. 

($\mathcal{E}_8$)
\begin{eqnarray}
p(c,d)=\mbox{Tr}(\mathcal{D}^{(2)}_d U^{(2)}_{-y}\mathcal{C}^{(1)}_c\rho_0\mathcal{C}^{(1)}_c U^{(2)\dagger}_{-y}),
\label{pcd1}
\end{eqnarray}   
where $c,d=+1,-1$. $\Rightarrow \expec{C(t^A)D_1(t^B)}=\sum_{c,d}p(c,d)cd=0$ where $p(+1,+1)=p(-1,-1)=p(+1,-1)=p(-1,+1)=1/4$. These subensembles give $\expec{I_Q}=0$.

\textit{A more efficient quantum key distribution (QKD) protocol}: In the above space-time Bell-CHSH test, Alice and Bob use subensembles $\mathcal{E}_5$ to $\mathcal{E}_8$ along with a small portion of subensembles $\mathcal{E}_1$ to $\mathcal{E}_4$, to test for eavesdropping/noise in the quantum channel. Remaining large portion of the subensembles $\mathcal{E}_1$ to $\mathcal{E}_4$ is used for secret key bit generation. Note that to separate the subensembles $\mathcal{E}_1$ to $\mathcal{E}_4$ from $\mathcal{E}_5$ to $\mathcal{E}_8$, they need to publicly announce only their random sequence of choice of observables, and Bob's random sequence of choice of $U_{+y},U_{-y}$, but not their measurement outcomes. They test for eavesdropping via following two steps: (I) They publicly announce their measurement outcomes corresponding to subensembles $\mathcal{E}_5$ to $\mathcal{E}_8$ and calculate $\expec{I_Q}$. If $|\expec{I_Q}|\approxeq 0$, then it is very likely that there was no eavesdropping. However, there is small probability that they obtain $|\expec{I_Q}|\approxeq 0$ even when there is eavesdropping as follows: If there is eavesdropping then $(\outpr{S_0}{S_0})^{\otimes N}$ will be transformed to $\rho_C=\rho_A\otimes\rho_B$. On $\rho_C$ Alice and Bob carry out local measurements. We know that $-2\le\expec{I_Q}_{\rho_C}\le 2$ ($\because\rho_C$ is separable). If we assume that all values of $\expec{I_Q}_{\rho_C}$ between $-2$ and $2$ are equally likely, then there is small probability that Alice and Bob obtain $|\expec{I_Q}_{\rho_C}|\approxeq 0$ even when there is eavesdropping. (II) To rule out this possibility, they publicly announce a few set of measurement outcomes chosen randomly from the subensembles $\mathcal{E}_1$ to $\mathcal{E}_4$, and look for their perfect correlation ($a=b,c=b,c=d$) and perfect anticorrelation ($a=-d$). Perfect correlation/anticorrelation in each set of measurement outcomes is possible if and only if particles were maximally entangled in each set (which implies no eavesdropping). They can also look for $\expec{I_Q}=4$ as it do not require an ensemble ($\because\expec{I_Q}=I_Q$ (Eq. (\ref{I_Q})), hence four set of measurement outcomes are sufficient to calculate $\expec{I_Q}$), unlike in Ekert and Wigner protocols (Table (\ref{Resorce_distrbtn})). If they obtain perfect correlation/anticorrelation in more than certain number of sets of measurement outcomes (this is to account for noise in the channel), then they can safely conclude that there was no eavesdropping, and generate the key bits. Else they have to discard the keys and start afresh. However, in the I-step itself if they obtain $|\expec{I_Q}|\gg 0$, then they can directly discard the keys i.e., no need to go for II-step.

If there was no eavesdropping, then they can generate secret key bits using the remaining large portion of subensembles $\mathcal{E}_1$ to $\mathcal{E}_4$ (whose outcomes are not publicly announced) as follows: Bob knows whether $B,D$ has been measured in the context of $A$ or $C$. Further $a=b,c=b,c=d$ (perfectly correlated). Hence both Alice's and Bob's measurement outcomes will be either $+1$ or $-1$. Hence they directly obtain the key. Where as $a=-d$ (perfectly anticorrelated). Hence if Alice's outcome is $\pm 1$, then Bob's outcome will be $\mp 1$. Hence one of them has to invert to obtain the key.

A comparison of our QKD protocol (ST) with other QKD protocols is given in Table (\ref{Resorce_distrbtn}). 
Instead of simply discarding approximately half of the measurements (corresponding to $\mathcal{E}_5$-$\mathcal{E}_8$), we are utilising it to test for eavesdropping, which is unlike in $\mbox{BB84}'$ and Ekert. This enhances the security further. Also ST has more keys than Ekert and Wigner, and approximately same as that of $\mbox{BB84}'$ (Table (\ref{Resorce_distrbtn})). Hence our protocol is more efficient than other three QKD protocols. Finally we note that, if Bob can store his qubit states in quantum memory till Alice publicly announces her random sequence of choice of measurement observables, then each of the $N$ measurements will be on correct context dependent unitarily evolved states. Then approximately full resource can be used for key generation i.e., wastage is close to zero. 

\begin{table}
\centering
\begin{tabular}{|c|c|c|c|c|}
\hline   & Key & Test(T)  & Discard(D) & Waste=T+D   \\ 
\hline  Ekert&   2/9&4/9 (E)&  3/9&  7/9 \\ 
\hline  Wigner&  1/4&3/4 (E)&  0&  3/4\\ 
\hline  ST(ours) &  1/2-$\epsilon$&1/2(E)+$\epsilon$&  0& 1/2+$\epsilon$   \\
\hline  $\mbox{BB84}'$ &  x/2&(1-x)/2 (F)&  1/2& (2-x)/2   \\ 
\hline 
\end{tabular} 
\caption{Fraction of the total resource distributed for various purposes \cite{QKD_expt_entphtn_Zeilinger}. First three QKD protocols are based on Bell's theorem. $\mbox{BB84}'$:=modified BB84 with entangled photons, E:=requires an ensemble for testing eavesdropping, F := requires (1-x)/2 of full resource for testing eavesdropping. 0$\ll$ x$<$ 1 i.e., x is close to one. $\epsilon(>0)$ is close to zero. Note that if Bob can store his qubit states in quantum memory till Alice publicly announces her random sequence of choice of measurement observables, then all entries in Discard column can be made zero. Consequently, more key bits can be generated in Ekert, ST, and $\mbox{BB84}'$. But storing quantum states against decoherence is a great challenge.}
\label{Resorce_distrbtn}
\end{table}

\textit{Conclusion}: We showed that if we insert context dependent unitary evolutions (via post selection) into normal Bell test, then it is possible to violate space-time Bell-CHSH inequality maximally. This does not contradict Tsirelson bound, as the latter does not take into consideration unitary evolutions and/or post selection. Further we showed that this leads to a more efficient quantum key distribution protocol. 

\textit{Acknowledgements}: I acknowledge discussions with Prof. T S Mahesh, Prof. R Srikanth, Aravinda, Deepak Khurana, Anjusha V S, and Soham Pal. 

\bibliographystyle{apsrev4-1}
\bibliography{bib_ch}

\begin{thebibliography}{11}
\bibitem [{IC2()}]{IC2}%
  \BibitemOpen
  \href@noop {} {}\bibinfo {note} {Instead if we define
  $I_1=A(t^A_i)B_1(t^B_k)+C(t^A_j)B_2(t^B_k)+C(t^A_j)D_1(t^B_l)-A(t^A_i)D_2(t^B_l)$
  and
  $I_2=A(t^A_i)B_2(t^B_k)+C(t^A_j)B_1(t^B_k)+C(t^A_j)D_2(t^B_l)-A(t^A_i)D_1(t^B_l)$,
  then also we obtain $I_C=\frac{1}{2}\times I_1+\frac{1}{2}\times
  I_2=(A(t^A_i)+C(t^A_j))B_m(t^B_k)+(C(t^A_j)-A(t^A_i))D_m(t^B_l)\le 2$ using
  non-contextual and dichotomic properties.}\BibitemShut {Stop}%
\end{thebibliography}

\appendix
\section{}
\subsection{Classical (noncontextual/local realistic) scenario without post selection}\label{appdx_no_postsel}
Charlie prepares a two particle state and sends one particle to Alice and the other to Bob. Alice measures observables (physical properties) $\mathbb{A},\mathbb{C}$ on her particle with probability $1/2,1/2$ respectively and Bob measures observables $\mathbb{B},\mathbb{D}$ on his particle with probability $1/2,1/2$ respectively. $\mathbb{A},\mathbb{C},\mathbb{B},\mathbb{D}$ are dichotomic observables which takes values $A,C,B,D(=\pm 1)$ respectively. Alice's and Bob's clocks are synchronised and their measurement events are space-like separated. Alice measures her $j^{\mbox{th}}$ particle state at time $t=t^A_j$. Immediately after Alice's measurement, Bob evolves his $j^{\mbox{th}}$ particle state under the Hamiltonian $H_k$ and then measures at time $t=t^B_j(>t^A_j)$, where $H_k$ is chosen randomly from the set $\{H_1,H_2\}$ with probability $\{1/2,1/2\}$ respectively, $j=1,2,...,N$. Bob knows each of $t^A_j$s. 

Observables $\mathbb{A},\mathbb{C},\mathbb{B},\mathbb{D}$ have preassigned (at the time of state preparation by Charlie) values $\{A(t^C),C(t^C),B(t^C),D(t^C)\}$ respectively and subsequent measurement by Alice and Bob, just reveals them. As they are causally disconnected, Alice's measurement outcome cannot influence Bob's measurement outcome, and hence $B(t^B_k),D(t^B_l)$ are independent of $A(t^A_i),C(t^A_j)$ (noncontextuality/locality). 
Hence $A(t^A_i),C(t^A_j),B_1(t^B_k),D_1(t^B_l),B_2(t^B_k),D_2(t^B_l)=\pm 1$, where $B_i,D_i$ implies that first there was evolution under Hamiltonian $H_i$, and then $B,D$ respectively was measured, $i=1,2$. Whole procedure is repeated $N$ times (each time Charlie follows same preparation procedure). Consider the quantity $I_1=A(t^A_i)B_1(t^B_k)+C(t^A_j)B_1(t^B_k)+C(t^A_j)D_1(t^B_l)-A(t^A_i)D_1(t^B_l)=(A(t^A_i)+C(t^A_j))B_1(t^B_k)+(C(t^A_j)-A(t^A_i))D_1(t^B_l)\le 2$ ($\because$ using context independence and dichotomic ($\pm 1$) properties. For a given ($j^{\mbox{th}}$) pair of particles, the values $A,C,B,D$ are fixed independent of measurement. $A,C$ can be measured simultaneously. Hence in $I_C$ we have both $A,C$. Similarly $B,D$). Similarly $I_2=A(t^A_i)B_2(t^B_k)+C(t^A_j)B_2(t^B_k)+C(t^A_j)D_2(t^B_l)-A(t^A_i)D_2(t^B_l)=(A(t^A_i)+C(t^A_j))B_2(t^B_k)+(C(t^A_j)-A(t^A_i))D_2(t^B_l)\le 2$. Hence $I_C=\frac{1}{2}\times I_1+\frac{1}{2}\times I_2=(A(t^A_i)+C(t^A_j))B_m(t^B_k)+(C(t^A_j)-A(t^A_i))D_m(t^B_l)$ where, $B_m(t^B_k)=(B_1(t^B_k)+B_2(t^B_k))/2=-1,0,1,D_m(t^B_l)=(D_1(t^B_l)+D_2(t^B_l))/2=-1,0,1$. One can easily verify that $-2\le I_C\le 2$. Another definition of $I_1,I_2$ also gives same bounds for $I_C$ \cite{IC2}. After the experiment, Alice and Bob meet and calculate the average value of $I_C$. It has the following upper bound: 
\begin{widetext}
\begin{eqnarray}
\expec{I_C}=\sum_{A(t^A_i),C(t^A_j)=-1,+1,B_m(t^B_k),D_m(t^B_l)=-1,0,+1}p(A(t^A_i),C(t^A_j),B_m(t^B_k),D_m(t^B_l))\nonumber\\ 
\times ((A(t^A_i)+C(t^A_j))B_m(t^B_k)+(C(t^A_j)-A(t^A_i))D_m(t^B_l))\le 2
\end{eqnarray}
\end{widetext}
where the joint probability $p(A(t^A_i),C(t^A_j),B_m(t^B_k),D_m(t^B_l))$ depends on initial state preparation by Charlie \cite{quant_info_neilson_chuang}. But we also have $\expec{I_C}=\expec{A(t^A_i)B_m(t^B_k)}+\expec{C(t^A_j)B_m(t^B_k)}+\expec{C(t^A_j)D_m(t^B_l)}-\expec{A(t^A_i)D_m(t^B_l)}$ \cite{IC1}. 
Hence we obtain the space-time Bell-CHSH inequality
\begin{eqnarray}
\expec{I_C}=\expec{A(t^A_i)B_m(t^B_k)}+\expec{C(t^A_j)B_m(t^B_k)}+\nonumber\\\expec{C(t^A_j)D_m(t^B_l)}-\expec{A(t^A_i)D_m(t^B_l)}\le 2.
\label{BellineqC}
\end{eqnarray}

\subsection{Classical (noncontextual/local realistic) scenario with post selection}
Initial setting is same as that in `classical scenario without post selection' . 
After post selection (similar to that in quantum scenario) we obtain $I_C=A(t^A_i)B_1(t^B_k)+C(t^A_j)B_2(t^B_k)+C(t^A_j)D_1(t^B_l)-A(t^A_i)D_2(t^B_l)=A(t^A_i)(B_1(t^B_k)-D_2(t^B_l))+C(t^A_j)(B_2(t^B_k)+D_1(t^B_l))$ ($\because$ of context independence i.e., locality). Due to locality and dichotomicity we also have $A(t^A_i),C(t^A_j),B_1(t^B_k),D_1(t^B_l),B_2(t^B_k),D_2(t^B_l)=\pm 1$. $\Rightarrow I_C=-4,-2,0,2,4$. Now consider the case where Charlie prepares the initial two particle state such that both outcomes $+1,-1$ corresponding to the observables $\mathbb{A},\mathbb{C},\mathbb{B},\mathbb{D}$ are equally likely (this is also the situation in corresponding quantum scenario and also in spatial Bell-CHSH inequality). Then even if Bob evolves such that $B_1(t^B_k)=B_2(t^B_k)=D_1(t^B_l)=1$ and $D_2(t^B_l)=-1$ always, still Alice and Bob obtain $\expec{I_C}=\expec{A(t^A_i)\times 2+C(t^A_j)\times 2}=2(\expec{A(t^A_i)}+\expec{C(t^A_j)})=2(0+0)=0$. But we also have $\expec{I_C}=\expec{A(t^A_i)B_1(t^B_k)}+\expec{C(t^A_j)B_2(t^B_k)}+\expec{C(t^A_j)D_1(t^B_l)}-\expec{A(t^A_i)D_2(t^B_l)}$. Hence $\expec{I_C}=\expec{A(t^A_i)B_1(t^B_k)}+\expec{C(t^A_j)B_2(t^B_k)}+\expec{C(t^A_j)D_1(t^B_l)}-\expec{A(t^A_i)D_2(t^B_l)}=0$.

However we also have 
\begin{widetext}
\begin{eqnarray}
\expec{I_{C}}=\sum\limits_{A(t^A_i),C(t^A_j),B_1(t^B_k),D_1(t^B_l),B_2(t^B_k),D_2(t^B_l)=-1,+1}p(A(t^A_i),C(t^A_j),B_1(t^B_k),D_1(t^B_l),B_2(t^B_k),D_2(t^B_l))I_C\nonumber\\
\le\sum\limits_{A(t^A_i),C(t^A_j),B_1(t^B_k),D_1(t^B_l),B_2(t^B_k),D_2(t^B_l)=-1,+1}p(A(t^A_i),C(t^A_j),B_1(t^B_k),D_1(t^B_l),B_2(t^B_k),D_2(t^B_l))4=4
\end{eqnarray}
\end{widetext}
where $p(A(t^A_i),C(t^A_j),B_1(t^B_k),D_1(t^B_l),B_2(t^B_k),D_2(t^B_l))$ is the corresponding joint probability. Hence $\expec{I_{C}}=\expec{A(t^A_i)B_1(t^B_k)}+\expec{C(t^A_j)B_2(t^B_k)}+\expec{C(t^A_j)D_1(t^B_l)}-\expec{A(t^A_i)D_2(t^B_l)}\le 4$. $\expec{I_{C}}=4$ corresponds to the following trivial case: Charlie prepares the initial two particle state such that $A(t^A_i)=C(t^A_j)=1$ always. Then Bob evolves such that $B_1(t^B_k)=B_2(t^B_k)=D_1(t^B_l)=1$ and $D_2(t^B_l)=-1$ always. Moreover it is an artefact of introducing time evolution and post selection. Hence we simply neglect it and consider only $\expec{I_C}=0$ derived in previous paragraph.

\end{document}